\documentclass{aa}  

\renewcommand\makeLineNumber{}

\usepackage{txfonts}
\usepackage[dvipsnames]{xcolor}
\usepackage{comment}
\usepackage[hidelinks]{hyperref}

\usepackage{graphicx}   
\usepackage[export]{adjustbox}

\usepackage{xcolor}
\usepackage{soul}
\setstcolor{RawSienna}

\newcommand{\lr}[1]{{\color{red} #1}}
\def\gsim{\;\raise0.3ex\hbox{$>$\kern-0.75em\raise-1.1ex\hbox{$\sim$}}\;}

\def\lsim{\;\raise0.3ex\hbox{$<$\kern-0.75em\raise-1.1ex\hbox{$\sim$}}\;}
\def\gsim{\;\raise0.3ex\hbox{$>$\kern-0.75em\raise-1.1ex\hbox{$\sim$}}\;}
\begin{document}

\title{Fast and ``lossless'' propagation of relativistic electrons along magnetized non-thermal filaments in galaxy clusters and the Galactic Center region. 
}

\author{
Eugene~Churazov \inst{1,2} 
\and
Lawrence~Rudnick \inst{3}
\and
Ildar~I.~Khabibullin \inst{4,1,2} 
\and
Marisa~Brienza \inst{5} 
\and
Alex~Schekochihin \inst{6,7} 
\and
Dmitri~Uzdensky \inst{6} 
} 

\institute{
Max Planck Institute for Astrophysics, Karl-Schwarzschild-Str. 1, D-85741 Garching, Germany 
\and 
Space Research Institute (IKI), Profsoyuznaya 84/32, Moscow 117997, Russia
\and
Minnesota Institute for Astrophysics, University of Minnesota, 116 Church St SE, Minneapolis, MN 55455, USA
\and 
Universitäts-Sternwarte, Fakultät für Physik, Ludwig-Maximilians-Universität München, Scheinerstr.1, 81679 München, Germany
\and
Istituto Nazionale di Astrofisica (INAF) - Istituto di Radioastronomia (IRA), via Gobetti 101, 40129, Bologna, Italy
\and
Rudolf Peierls Centre for Theoretical Physics, Department of Physics, University of Oxford, Oxford, OX1 3PU, United Kingdom
\and 
Merton College, Oxford, OX1 4JD, United Kingdom
}

\abstract{
Relativistic leptons in galaxy clusters 
lose their energy via radiation (synchrotron and inverse Compton losses) and interactions with the ambient plasma. At $z\sim0$, pure radiative losses limit the lifetime of electrons emitting 
at $\sim {\rm GHz}$ frequencies to $t_{r}\lesssim 100 \,{\rm Myr}$.
Adiabatic losses can further lower 
Lorentz factors of electrons trapped in an expanding medium. 
If the propagation speed of electrons relative to the ambient weakly magnetized (plasma $\beta\sim10^2$) Intracluster Medium (ICM) is limited by the Alfvén speed, $\varv_{a,{\rm ICM}}=c_{s,{\rm ICM}}/\beta^{1/2}\sim 10^7\,{\rm cm\,s^{-1}}$, 
GHz-emitting electrons can travel only $l\sim \varv_{a,{\rm ICM}}t_r\sim 10\,{\rm kpc}$ relative to the underlying plasma. Yet, elongated structures spanning hundreds of kpc or even a few Mpc are observed, requiring either a re-acceleration mechanism or another form of synchronization, e.g., by a large-scale shock. 
We argue that filaments with ordered magnetic fields supported by non-thermal pressure have $\varv_{a}\gg \varv_{a,{\rm ICM}}$ and so can provide such a synchronization even without re-acceleration or shocks. In particular, along quasi-stationary filaments, electrons can propagate without experiencing adiabatic losses, and their velocity is not limited by the Alfvén or sound speeds of the ambient thermal plasma.
This model predicts that along filaments that span significant pressure gradients, e.g., in the cores of galaxy clusters,
the synchrotron break frequency $\nu_b\propto B$ should scale with the ambient gas pressure as $P^{1/2}$. The emission from such filaments should be strongly polarized due to the magnetic field being ordered along these structures. 
While some of these structures can be observed as ``filaments'', i.e., long and narrow bright structures, others can be unresolved and have a collective appearance of a diffuse structure, or be too faint to be detected, while still providing channels for electrons' propagation.
We examine several cases of visible filamentary structures in tailed radio galaxies and in a cluster relic, where "lossless" propagation provides an attractive alternative to other mechanisms for explaining the observed spectral behaviors. 
}

\titlerunning{Fast \& furious}

  
\maketitle

\section{Introduction}

\label{s:intro}
\begin{figure}
\centering
\includegraphics[trim=4.6cm 1.9cm 4.6cm 1.8cm,clip,width=0.4895\columnwidth]{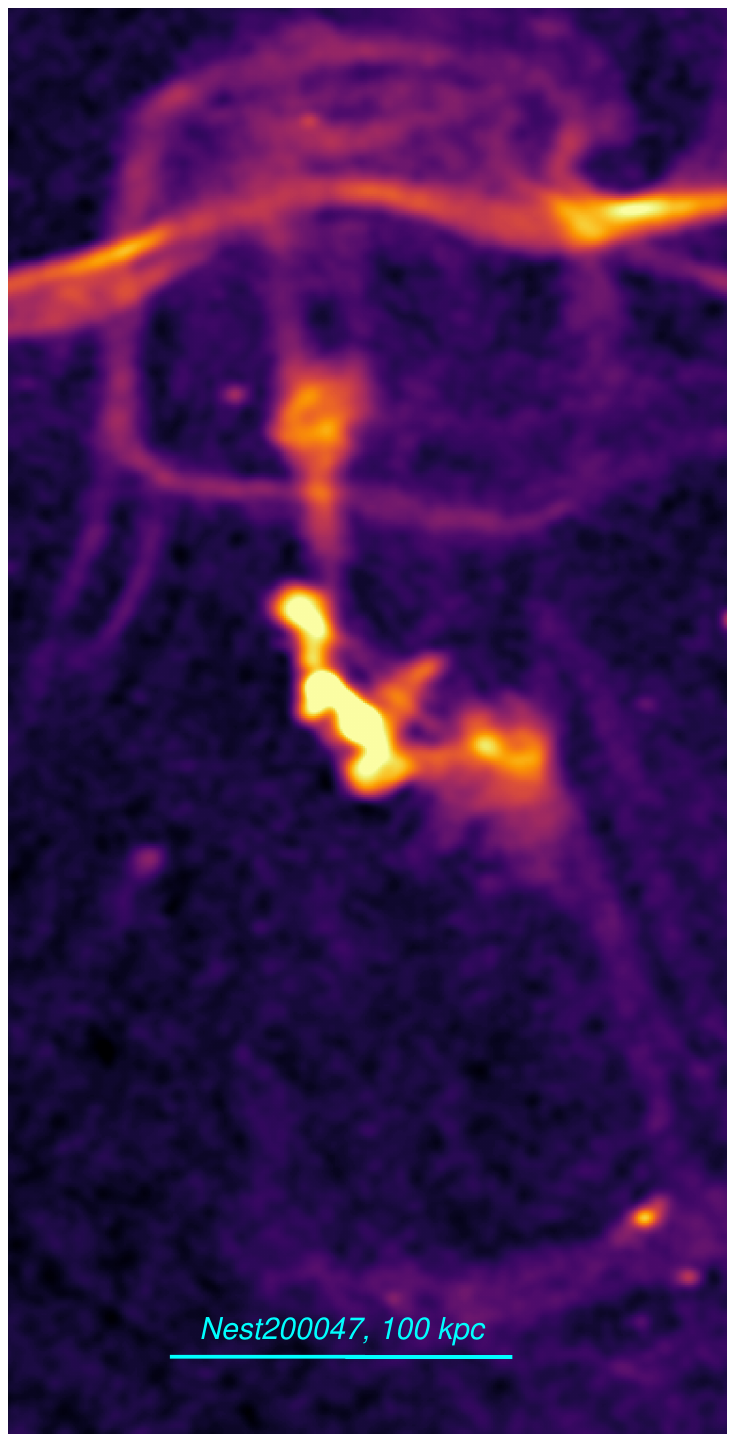} 
\includegraphics[angle=0,trim=5cm 1.9cm 5cm 1.8cm,clip,width=0.458\columnwidth]{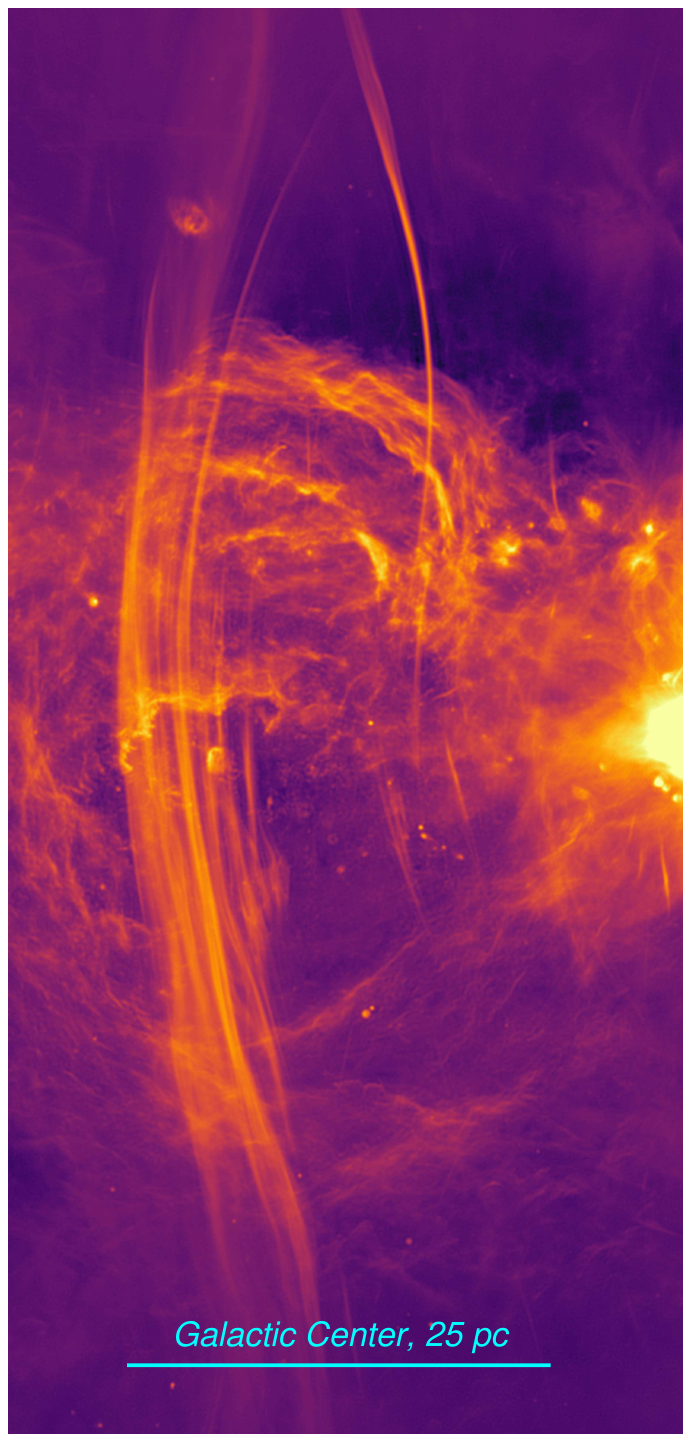} 
\includegraphics[angle=0,trim=0cm 0cm 0cm 1.5cm,clip,width=0.95\columnwidth]{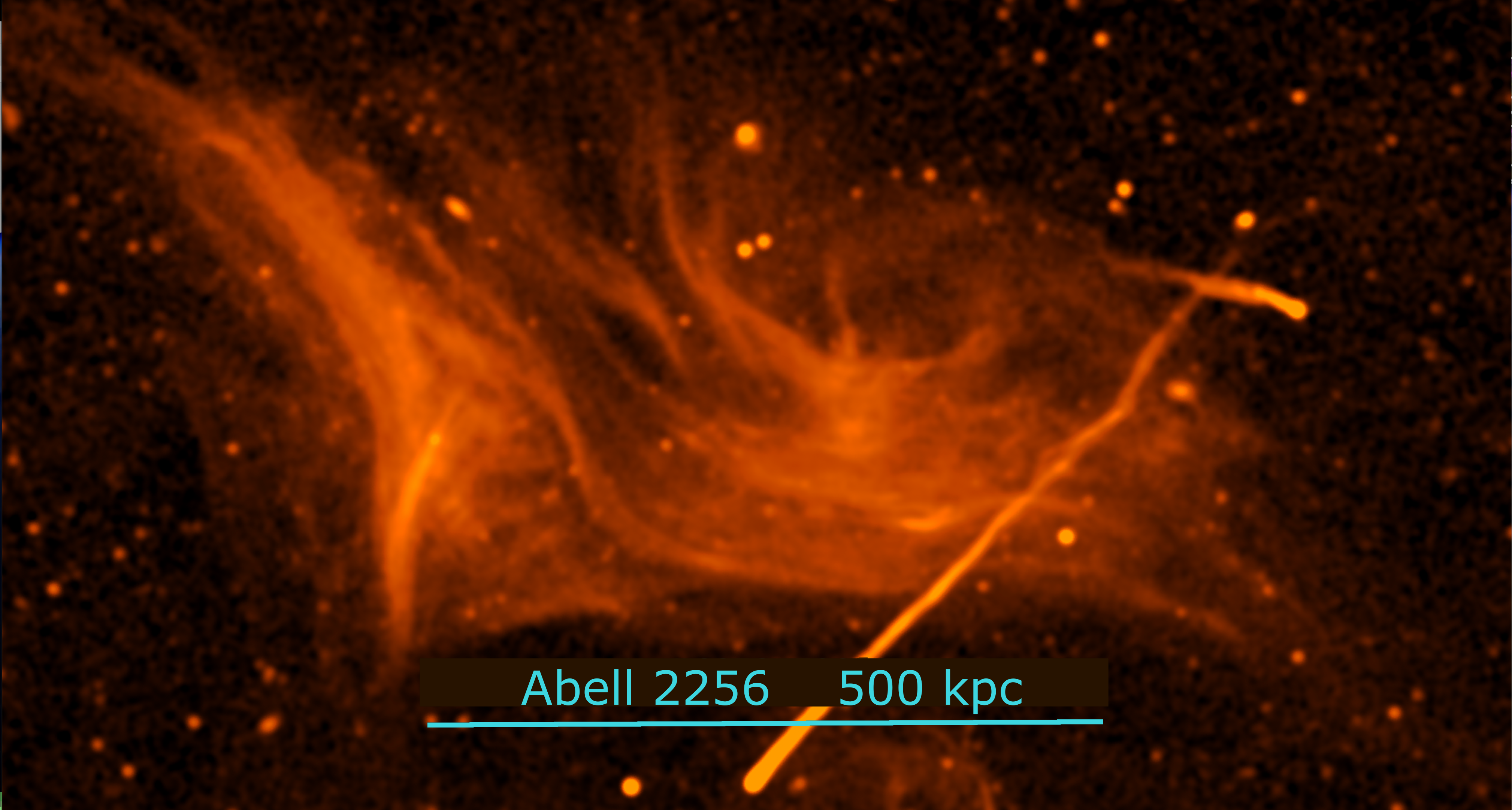} 
\includegraphics[angle=0,trim=0cm 0.5cm 0cm 0cm,clip,width=0.95\columnwidth]{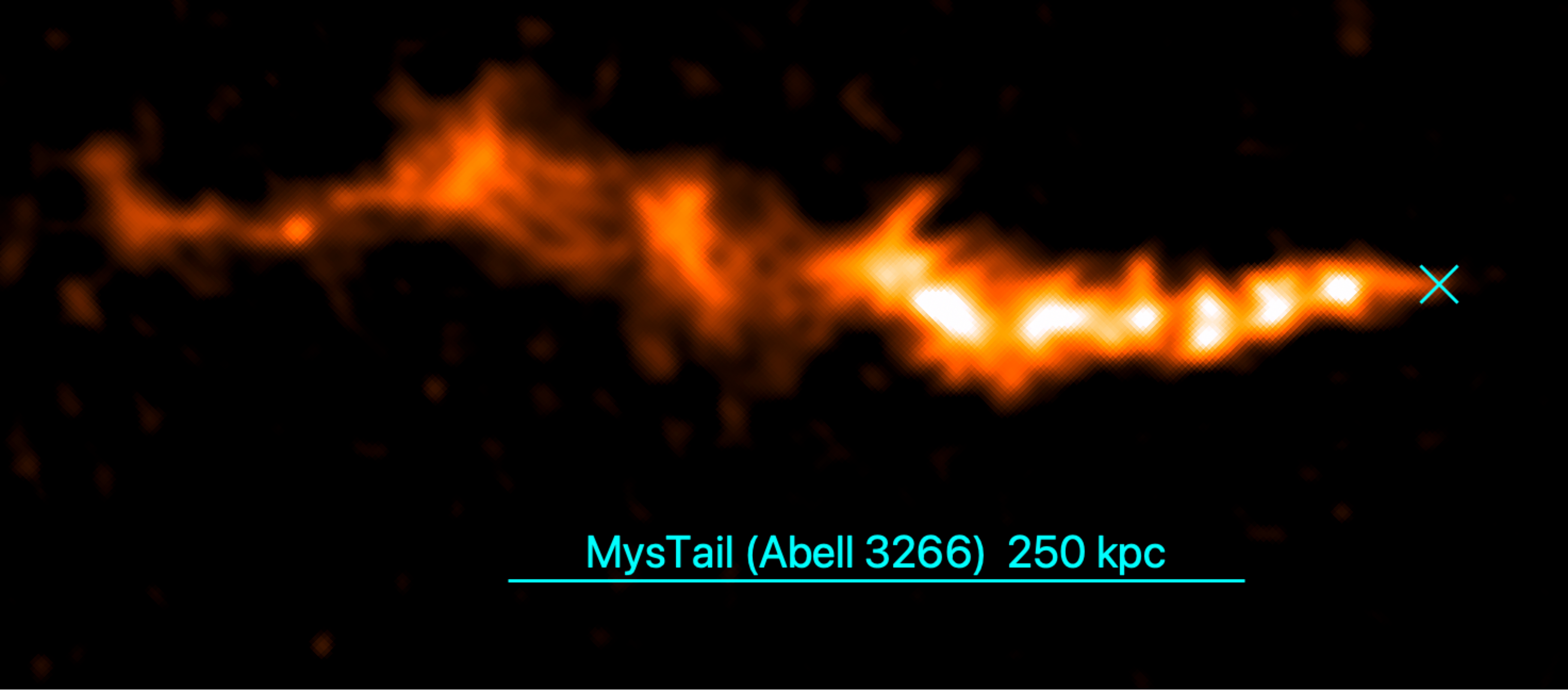} 
\caption{Examples of radio-bright filaments in clusters and the Galactic Center, where fast propagation of relativistic electrons might be important. 
Top panel;  the core of Nest200047 group, (left, \citealt{2021NatAs...5.1261B}, LOFAR image at 144~MHz) and the Galactic Center Non-thermal filaments (right, \citealt{2019Natur.573..235H}, MeerKAT image at 1.3~GHz).\lr{} Middle panel: the peripheral radio relic at the edge of the X-ray halo in Abell 2256 (\citealt{2014ApJ...794...24O}, VLA image at 1.5~GHz).  In all cases, the filaments are long, comparable to the characteristic size of the system. They appear ``laminar", and from the polarization measurements in the GC and Abell 2256,  the magnetic fields are oriented along the filaments 
\citep[e.g.,][]{2019ApJ...884..170P,2006AJ....131.2900C}. In the case of the Galactic Center, we also know that the field is dynamically strong \citep[e.g.,][]{2006JPhCS..54....1M}. Bottom: the tailed radio galaxy, MysTail in Abell 3266, showing three parallel 40-50 kpc long filaments embedded in the tail. The host position is marked with an "X". 
}
\label{f:img}
\end{figure}

There are many examples of synchrotron-powered radio sources that have complicated filamentary structures (see Fig.~\ref{f:img}), which are becoming more and more common as high-sensitivity radio observations become routinely available. Those include the Galactic Center Non-thermal Filaments \citep[NTF, e.g.,][]{1984Natur.310..557Y}, filaments in the cores of galaxy groups and clusters \citep[see, e.g.,][for examples of filamentary structures in the radio lobes of M87]{2000ApJ...543..611O} and radio relics in the outskirts of galaxy clusters \citep[see][for a review]{2019SSRv..215...16V}, and tails of radiogalaxies \citep[e.g.,][]{2024A&A...692A..12V}.

Diffusive shock acceleration (DSA) mechanism \citep[e.g.,][]{1977DoSSR.234.1306K,1978ApJ...221L..29B,1978MNRAS.182..147B}, predicts a power-law spectrum of accelerated relativistic particles  $dN/d\gamma\propto \gamma^{-p}$ up to very high energies, and is plausibly responsible for the production of relativistic electrons in these objects.   
In addition to acceleration by large-scale shocks, AGNs release accelerated relativistic particles into the ICM, either through directed flows or escape from old radio galaxy structures.
Without re-acceleration, these electrons should lose their energy due to synchrotron radiation, inverse Compton Scattering (IC), adiabatic losses, and interaction with ambient plasma. Since the radiative losses depend strongly on the particle energy ($\propto \gamma^2$), a high-energy cutoff develops in the particle spectrum, which reveals itself in the observed synchrotron spectra. This break can be used to infer the ``age'' of the particle distribution if the magnetic field is known and no additional re-acceleration is taking place. The adiabatic losses scale linearly with $\gamma$ and, therefore, do not introduce a break in the particle spectra but merely shift the existing energy break (if any). While  ``aged'' spectra are commonly seen, there are cases when the inferred age is shorter than the estimated dynamic age of the source or the time scale for electrons' streaming along the magnetic field with velocities lower than the Alfvén speed in ambient thermal plasma \cite[e.g.,][]{2017PhPl...24e5402Z}. While various explanations for these discrepancies have been suggested (including a possibility of re-acceleration, e.g., \citealt{1976ApJ...203..313P,2001MNRAS.320..365B,2001ApJ...557..560P}), we discuss below a set of conditions that might simultaneously lead to the minimization of losses and fast propagation of electrons along filamentary structures. 

Let us provide a few examples of objects featuring long and narrow filaments of synchrotron emission, and where the simplest estimates of the age and losses often appear in tension with observations. 

The first group of such objects includes rising bubbles and filaments of relativistic plasma in the cores of galaxy groups and clusters, 
e.g., M87 \citep{2000ApJ...543..611O}, Nest200047 \citep[][see Fig.~\ref{f:img}.]{2021NatAs...5.1261B,2025A&A...696A.239B} and Abell~194 \citep{2022ApJ...935..168R}, to name a few. Typical sizes of these radio-bright structures are in the range from tens to a couple of hundred kpc (the most prominent structure in the Nest200047 has a projected length of $\sim 350\,{\rm kpc}$).
Their age estimates can be based on the dynamics of buoyant bubbles \citep[e.g.,][]{1973Natur.244...80G,2000A&A...356..788C,2018MNRAS.478.4785Z}, which move down the pressure gradient of approximately hydrostatic atmospheres with velocities  $\lesssim c_{s,{\rm ICM}}$. For instance, in the Nest200047 group \citep{2025A&A...696A.239B}, the estimated dynamic age of some of the structures is $t\sim 200\,{\rm kpc}/500\,{\rm km\,s^{-1}}\sim 4\times 10^{8}\,{\rm yr}$. 
Fitting the shape of the observed radio-band spectra in selected regions  (assuming magnetic fields $\sim 2\,{\rm \mu G}$, see Sect.~\ref{s:radiative}) yields a factor of $\sim2$ smaller age when pure radiative losses are considered. The factor of 2 does not look irrefutable, but once the adiabatic losses are added and the magnetic field is set to values comparable to the gas thermal pressure, the discrepancy reaches almost an order of magnitude.  

Another group of objects that might be affected by fast propagation of electrons includes very long tails of radio galaxies, e.g., 3C129 \citep{1973A&A....26..423J,1976ApJ...203..313P}, Tail C in Abell~2256 \citep{2014ApJ...794...24O}, and the tail in the south of A1033 \citep{2017SciA....3E1634D}. There, a prominent steepening of the spectrum is very clear in the vicinity of the AGN, but the spectrum flattens and stays relatively constant towards the end of the tail. A long tail in Abell 3266, on the periphery of the X-ray emission, shown in Fig. \ref{f:img}, displays similar behaviors.

The third group consists of radio relics \citep[see, e.g.,][for a review]{2019SSRv..215...16V}.  One of the best and earliest known relics showing long, prominent filamentary structures is Abell 2256 \citep{1976A&A....52..107B}, also shown in Fig.\ref{f:img}. 
These structures are very likely created by merger shocks, which accelerate and/or re-accelerate relativistic particles. Relics have integrated radio spectra with the spectral index approaching $-1$. Here, the spectral index $\alpha=\frac{d \ln{S_\nu}}{d\ln{\nu}}$, where $S_\nu\propto \nu^\alpha$ is the emission spectral energy density. In the context considered here, $\alpha$ is always negative, consistent with an optically thin plasma emitting synchrotron radiation.
At the same time, the modest values of shock Mach number inferred from X-ray surface brightness jumps suggest (in the DSA framework) steeper indices.  The most actively discussed scenario explains these relics' properties by strong dependence of electrons' acceleration efficiency on the shock Mach number \citep[e.g.,][]{2019ApJ...876...79K} coupled with variations of the shock Mach number on small scales seen in numerical simulations \citep[e.g.,][]{2021MNRAS.506..396W,2021MNRAS.500..795D,2024arXiv241111947W,2025ApJ...978..122L}. These variations can be associated with density or velocity fluctuations in the ICM or with multiple shocks \cite[][]{2022MNRAS.509.1160I}.   Some of these relics feature bright filamentary structures with relatively constant spectra along them. The ``Toothbrush'' relic presents another version of this puzzle:  there, the integrated spectra have strikingly similar slopes ($\Delta \alpha \approx 0.03$) along the entire structure spanning 1-2 Mpc  \citep{2020A&A...642L..13R}. 
If parts of the relic are interconnected by a network of filamentary structures, the fast propagation of electrons can contribute to the uniformity of the spectra. In this case, the spectrum slope may depend on the Mach number distribution in the entire structure.

The last example deals with the 10-100~pc-long Non-Thermal Filaments (NTF) in the Galactic Center, including the famous Radio Arc  \citep[e.g.,][see Fig.~\ref{f:img}]{1984Natur.310..557Y,2019Natur.573..235H}. In some of these filaments, which are possibly 
powered by Pulsar Wind Nebulae \citep[PWN, e.g.,][]{2002ApJ...581.1148W,2017SSRv..207..235B,2019MNRAS.484.4760B,2020ApJ...890L..18T,2024MNRAS.530..254Y}, large Lorentz factors of particles and a strong magnetic field produce synchrotron emission even in the X-ray band. The lifetime of such particles is short ($\sim 30\,{\rm yr}$ assuming dynamically important magnetic field), and the length of observed structures requires propagation velocity to be a sizable fraction of the speed of light \citep{2002ApJ...581.1148W,2020ApJ...893....3Z,2024A&A...686A..14C}. X-ray emission is strongly polarized, close to the maximal possible polarization of the synchrotron emission in a uniform field aligned with the direction of filaments. 
Some NTFs form a sequence of almost equi-distant parallel filaments \citep[e.g.,][]{2001ApJ...563..163L}. The length of individual filaments decreases monotonically in the direction perpendicular to filaments, forming an overall shape resembling a triangle (sometimes called a ``harp"); see, e.g., Fig.8 in \cite{2022ApJ...925..165H} for several examples. The central part of the Radio Arc in Fig.~\ref{f:img} also features a triangle-shaped extension that points towards a PWN. In the model of \cite{2020ApJ...890L..18T}, these structures are due to the propagation of leptons (the leading front) with the Alfvén speed $\varv_a\sim 40\,{\rm km\,s^{-1}}$, while the source of the leptons is at the tip of the triangle. In this case, the shape of the ``triangle'' (half angle typically larger than 45 degrees) bears partial analogy\footnote{The analogy is partial because unlike isotropic sound waves, the leptons move along the field lines in the ambient volume.} to the Mach cone of a supersonic jet, suggesting that the source is moving at a velocity comparable to or lower than~$\varv_{a}$. This would be consistent with one of their models, where massive stars with velocities a few tens ${\rm km\,s^{-1}}$ produce leptons. However, the leptons might come from PWNs (our preferred model), which have an order of magnitude (or more) higher velocities and, therefore,  the propagation speed of leptons has to be proportionally higher -- at least a few 100 ${\rm km\,s^{-1}}$ independently of the details of the particle propagation mechanism. This value is much higher than the sound or Alfvén speeds in the ISM.  

Here, we shall assume that all filamentary objects in these diverse groups share the same properties -- a strong magnetic field aligned along the low-density filaments -- and argue that in these conditions, relativistic electrons can propagate very rapidly and experience minimal losses. While another mechanism, such as re-acceleration \citep[e.g.,][]{1976ApJ...203..313P,2001MNRAS.320..365B,2001ApJ...557..560P},
might be responsible for maintaining long-lived synchrotron emission in the bulk of the ICM, we discuss here situations where lossless propagation plays an attractive, plausible role.

Unless stated otherwise, we consider a filament of length~$L$ supported by a strong magnetic field~$B_f$. It is embedded in thermal plasma with gas pressure $P_{t}\sim B_f^2/8\pi$. Here and below,  quantities with subscripts $f$ and $t$ refer to the filament and ambient thermal plasma, respectively. We further assume that the filament contains relativistic particles 
and, possibly, some amount of thermal gas, with mass density much smaller than the density of the ambient thermal plasma, $\rho_f\ll \rho_t$. With these assumptions, $\beta_f\lesssim 1$ or, possibly, $\beta_f\ll 1$. In the latter case, the magnetic pressure dominates inside the filament, while in the former, it makes up half of the filament pressure. In all cases considered here, we assume that the total pressure $P_{\rm particles}+B^2/8\pi$ in the filament is approximately equal to the ambient medium (ICM) pressure. This might not be the case for overpressurized bubbles in the proximity of an AGN in cluster cores, which drive shocks in the ambient gas. Such objects are not considered here. ``Our'' filaments could be advected by the ICM motions and expand/contract 
to maintain an approximate pressure equilibrium with the ambient gas.     

\section{Losses}
\label{s:losses}
Ignoring Coulomb losses, which that are not dramatically important for particles with $\gamma\gtrsim 10^4$ \citep[e.g.,][]{1999ApJ...520..529S,2001ApJ...557..560P} of interest for this study\footnote{If filaments are devoid of thermal plasma, the Coulomb losses will be absent and, therefore, particles with lower Lorentz factors can survive for a long time. We return to this issue in Sect.~\ref{s:discussion}}., We briefly discuss below the radiative and adiabatic losses. 

\subsection{Adiabatic losses }
\label{s:adiabatic}

This section considers adiabatic losses of relativistic particles in isolation from all other possible losses. We first consider the case when particles are trapped in the tangled magnetic field of the 
slowly expanding Lagrangian fluid element with volume~$V$. We further assume that some level of scattering is present, so the pitch-angle distribution remains isotropic.
Such particles should suffer from adiabatic losses. The Lorentz factor of particles changes as $\gamma\propto V^{1-\Gamma}$, where $\Gamma=4/3$ in the ultrarelativistic limit. If there is a break in the electron distribution function at~$\gamma_b$, it will move down accordingly. We further assume that the energy density of the magnetic field also goes down with expansion, e.g., for a tangled field, $B_f\propto V^{-2/3}$. 
With these assumptions, the break frequency in the observed synchrotron spectrum evolves as $\nu_b\propto \gamma_b^2B_f \propto V^{-4/3}$.  This is relevant, for example,  for buoyant bubbles of relativistic plasma in cluster cores \citep{2001ApJ...554..261C}. If the bubbles are in pressure equilibrium with the ambient gas, then their volume $V\propto P_{t}^{-1/\Gamma}$, where $\Gamma$ is again $4/3$, so $\nu_b\propto P_{t}$. Accordingly, as the bubbles move down the pressure gradient, the decrease of the break frequency should follow the drop of pressure (or faster if radiative losses are important).    

However, if there is a pre-existing filament (a ``channel") that is stationary,  then in the test-particle regime, there are no adiabatic losses for a particle propagating along the channel. This remains valid independently of the presence or absence of pitch-angle scatterings by spatially fluctuating magnetic field or changes 
of the medium's energy density (and hence pressure) along the filament. This is also true if the filament is wider at one end and narrower at the other. What matters is that the scattering centers, on average, do not move away from the particle. This configuration closely resembles the model of tailed radio sources by \cite{1973A&A....26..423J}, where the tail associated with a high-speed galaxy is replaced with a filament. In that model, the adiabatic losses are explicitly neglected \citep[see also][for a different model]{1976ApJ...203..313P}. In application to our hypothetical filaments, we note that the amount of adiabatic losses experienced by particles during a single journey along the filament can be linked to the change in the filament volume during that journey. Therefore, if the particle can propagate fast enough (so that the filament volume only changes by a small amount), the changes in the particle's Lorentz factor along the filament associated with adiabatic expansion will be small, too (even though all particles do lose energy on the filament-expansion time scale). \footnote{Observationally  $\nu_b = \gamma^2_b B_f$ can change depending on the local value of $B_f$.} In other words, the adiabatic losses are inevitable, but they are not necessarily reflected in the shape of the electron spectrum along the filament.

\subsection{Synchrotron and IC losses}
\label{s:radiative}
These losses are set by the energy density of the magnetic and radiation fields\footnote{We assume that photons composing the radiation field are isotropic and have sufficiently small energy so that Klein-Nishina cutoff is not relevant for the energy exchange with relativistic electrons.} 
\begin{equation}
    \dot{\gamma}=-\gamma^2 \frac{4}{3}\sigma_T c\left ( U_{\rm rad}+\frac{B^2}{8\pi} \right),
\end{equation}
assuming the pitch angle isotropization is faster than the cooling losses, where $\sigma_T$ is Thomson cross section, $c$ is the speed of light, $U_{\rm rad}$ is the energy density of radiation. Unlike the adiabatic losses that vanish for a stationary filament (Sect.~\ref{s:adiabatic}), radiative losses are inevitable.

The longest lifetime of particles emitting at a given frequency is achieved when 
\begin{equation}
    B=B_1=\left ( \frac{8\pi}{3} U_{\rm rad}\right )^{1/2},
    \label{e:b1}
\end{equation}
see \cite{1969A&A.....3..468V}. If the radiation energy density is dominated by CMB photons, then the upper limit on the particle lifetime is  
\begin{equation}
    t_{\rm max}\sim 1.54\times 10^8 \,\nu_{c,{\,\rm GHz}}^{-1/2} (1+z)^{-3} \,{\rm yr},
    \label{e:tmax}
\end{equation}
where $z$ is the redshift and $\nu_c$ can be derived from the observed spectra\footnote{Strictly speaking, the geometry of the magnetic field has to be accounted for.}. For $z=0$, $B_1\sim 1.9\,{\rm \mu G}$, and for weaker or stronger $B$, the lifetime will be shorter. This argument is often used to place an upper limit on the age of a synchrotron source or on the distance $d\lesssim\varv \,t_{\rm max}$ that electrons can travel with velocity~$\varv$.

The only way to reduce the aging of the electron population with the distance along the filament from the acceleration site is to increase the propagation velocity of electrons (or invoke a re-acceleration mechanism, e.g.,  \citealt{1976ApJ...203..313P,2001MNRAS.320..365B}). In particular, the assumption that particle velocities are limited by the Alfven speed $c_{s,t}/\beta_t^{1/2}$ in thermal plasma ($\beta_t\sim 10^2$) implies that at GHz frequencies the aging is important on scales $\sim 10\,{\rm kpc}$ and up to $\sim 100\,{\rm kpc}$ if the electrons are advected with trans-sonic velocities. 

Since we consider here a ``strong-field'' case, i.e., $B_f^2/8\pi\sim P_{t}$, it is interesting to note that near the virial radius of present-day clusters, the corresponding value of $B_f$ would be $\sim {\rm \mu G}$ and, therefore, the energy density of the magnetic field in such filaments will be comparable to the CMB energy density. These conditions maximize the lifetime of particles emitting at a given frequency, implying that objects like radio relics might be visible for a long time \citep{2023A&A...670A.156C}. In contrast, in the cluster cores, the ``strong-field'' condition corresponds to $B_f\sim 20-50\,{\rm \mu G}$, synchrotron losses dominate particles' aging, and the lifetime of particles emitting at a given frequency is relatively short. 

\begin{figure}
\centering
\includegraphics[angle=0,clip,trim=0.5cm 5.cm 1cm 3cm,width=0.99\columnwidth]{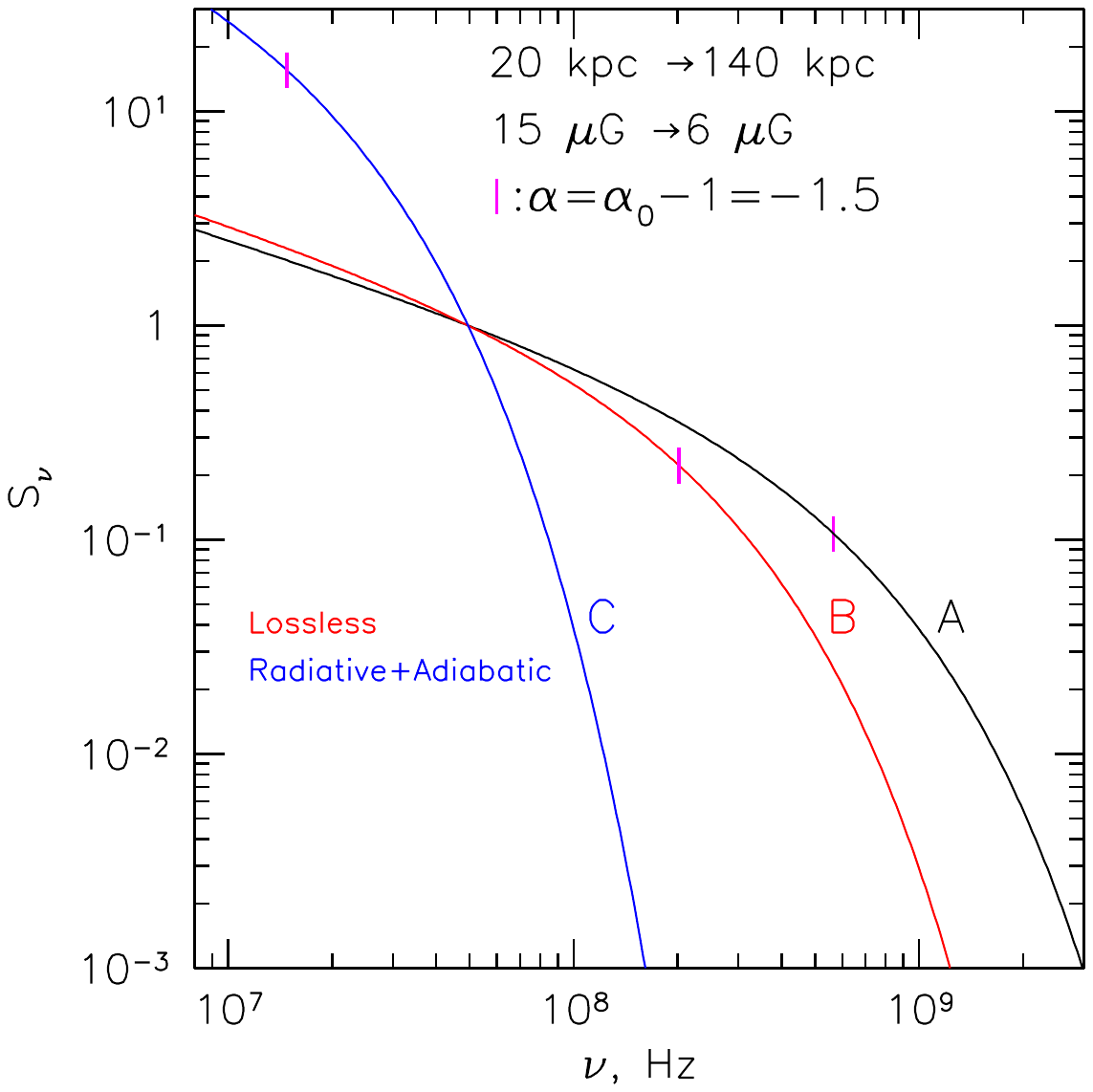}
\caption{
Expected steepening of synchrotron spectra as particles move from the cluster core in the radial direction from 20 to 140 kpc. All spectra are normalized to unity at $\nu=5\times 10^7\,{\rm GHz}$. The magnetic energy density is assumed to follow the ICM pressure profile $P_t(r)$, i.e., $B_f^2/8\pi=P_t(r)$. For these simulations, we adopted a simple power law pressure profile  $P_t\approx 2\times 10^{-10} (r/{\rm kpc})^{-1}\,{\rm erg\, cm^{-3}}$, derived from X-ray observation of the NEST200047 group \citep{2025A&A...699A.375M}. The solid back curve (marked with ``A'') shows the synchrotron spectrum at the initial position at $r_1=20\,{\rm kpc}$. It has a low-frequency slope $\alpha_0=-0.5$ and is ``aged'' in the $15\,\mu{\rm G}$ field for $3\times 10^7\,{\rm yr}$ so that a break develops in the spectrum. The spectra marked with ``B'' and ``C'' show the evolved spectra at the final position at $r_2=140\,{\rm kpc}$. These spectra represent two extreme limits. Namely, for ``C'', the electrons are moving ``in a bubble'' from $r_1$ to $r_2$ with the velocity $\varv=500\,{\rm km\,s^{-1}}$ (just below the sound speed in the group ICM $c_s\sim700 \,{\rm km\,s^{-1}}$) and suffer from the adiabatic and radiative losses. In addition to the evolution of the particle spectrum, the magnetic field is lower at $r_2$. All these effects combined lead to a very steep spectrum at the final position. In contrast, for the spectrum ``B'', we assume that electrons quickly propagate ``along a static filament'' and do not suffer from any losses. The only reason why the spectrum ``B'' is steeper than ``A'' is that the magnetic field is lower at the final position, hence a lower critical frequency $\nu_c\propto B$. For each curve, the position of the break frequency (defined here as a frequency where the spectral slope is $\alpha=\alpha_0-1$) is shown with a small vertical magenta bar. This plot illustrates that if the initial spectrum has a break around 1~GHz, and no re-acceleration is present, any ``subsonic'' regime of propagation (blue lines) will result in a very steep spectrum at 100~MHz. If ``fast-track'' routes are available for a fraction of electrons, this might reduce the apparent steepening dramatically. 
}
\label{f:age}
\end{figure}

Let us illustrate the expected magnitude of the combined adiabatic and radiative losses on the spectrum of a subsonically rising ``bubble'' in the core of the Nest200047 group in Fig.~\ref{f:age}. This object has a large linear size of AGN-powered radio structures  \citep{2021NatAs...5.1261B} and a steep radial pressure profile, while the thermal gas temperature is modest $\sim 2\,{\rm keV}$. This makes this object particularly suitable for getting the most stringent constraints on the electrons' losses associated with long rise time (radiative losses) and large pressure gradient (adiabatic losses). From this figure, it is clear that a combined action of adiabatic and radiative losses for $B_f^2/8\pi=P_t$ would lead to a very steep spectrum (the blue curve in Fig.~\ref{f:age}) at a distance of 140~kpc from the core. In \cite{2025A&A...696A.239B}, the lack of very strong steepening is discussed in terms of the re-acceleration scenario. Here we consider an alternative possibility -- low-beta filaments, which in the most optimistic scenario can eliminate these losses: see the red curve in Fig.~\ref{f:age}. Essentially, the blue and red curves correspond to the maximal- and minimal-loss scenarios, respectively, and the difference in the predicted spectra is large. In this study, we focus on the latter scenario. In this (minimal-loss) limit, the spectrum of electrons is the same over the entire filament. Therefore, in the observed synchrotron emission, the break frequency varies along the filament solely due to variations of $B\sin\theta$, where $\theta$ is the angle the magnetic field makes to the line of sight.

Yet another convincing example of a competition between radiative losses and rapid propagation comes from PWNs. X-rays from the ``misaligned'' filaments associated with Pulsar Wind Nebulae provide solid evidence that on scales $\sim 1-10\,{\rm pc}$, relativistic leptons (with $\gamma\sim 10^8$) propagate with the velocity of at least $\sim 10^4\,{\rm km\,s^{-1}}$ \citep[e.g.,][]{2008AIPC..983..171K,2023ApJ...950..177K,2024ApJ...976....4D}. This follows from the synchrotron cooling time estimates and the observed length of the filaments. The motion of the $\sim$ parsec-long X-ray-bright filament together with the PWN in the Guitar Nebula \citep[see Fig.~4 in][]{2022ApJ...939...70D} over 20 years corroborates this conclusion. In fact, if the length of these filaments is set not by cooling time but by the changes in the filaments' geometry/properties, the velocity should be even higher than estimated using the observed length and Eq.~(\ref{e:tmax}). 
For instance, the filament might transition to a high-$\beta$ type at its end, effectively preventing further fast propagation of particles by enhancing the scattering rates \citep[as in][]{2024MNRAS.532.2098E}. In this case, particles might be able to travel back and forth along the filaments, erasing gradients associated with radiative cooling along filaments. The equipartition fields in these objects are not dissimilar from the values that can be found in clusters (assuming $B^2/8\pi\sim P_t$). However, for electrons emitting at frequencies $\sim 1\, {\rm GHz}$, the same propagation velocities would translate into distances of $\sim 100\,{\rm kpc}$ over their longer radiative cooling time. 

\subsection{Interactions with plasma waves}
\label{s:waves}
Another process pertinent to the propagation of particles is the interaction with plasma waves, which can scatter relativistic particles (\citealt{1969ApJ...156..445K}; see, e.g., \citealt{2017PhPl...24e5402Z,2023A&ARv..31....4R} for reviews).  The condition of the strong field in the filament adopted here implies that external driving, e.g., turbulence in the ICM, is unlikely to affect the fields on scales comparable to gyroradii of relativistic electrons. In the test-particle regime, the relativistic electrons will stream \citep{1975MNRAS.172..557S} along the filament with a velocity on the order of $c$.  However, it is often assumed that streaming particles excite waves that, in turn, scatter these particles. In the absence of other damping mechanisms, the waves (and the particles' scattering rate) grow until the streaming velocity drops below the local Alfven speed. Such a process can dramatically reduce the cosmic rays' propagation speed away from the source in the ICM or ISM.

In a strong-field non-thermal filament, however, the Alfvén speed $\varv_{A,f}$ can be very high, much higher than the Alfvén speed $\varv_{A,t}$  in the ambient high-$\beta$ plasma or its sound speed $c_s$. In fact, $\varv_{A,f}$ can be as high as ${\rm O(1)}c$ if the filament is completely devoid of thermal gas. This has important implications. Since the filament itself is supposedly created (see Sect.~\ref{s:adiabatic}) on timescales of the order of, or longer than, the sound crossing time in the ambient plasma, this means that the gradient of relativistic particles' energy density along the filament will be small during the formation of the filament. The second implication is that, once the filament is formed, the gradients caused by any additional injection of fresh particles will be erased quite quickly or, if the streaming velocities fall below $\varv_{A,f}$, the streaming instability will not be triggered at all. In any case, a high  Alfvén speed implies that both adiabatic and radiative losses on the time scale of particles' propagation along the filament are small. Therefore, the situation where there is no difference in the ``effective age'' of the relativistic electrons' distribution along the filament arises naturally.

\section{Formation and ``survival'' of filaments}
\label{s:survival}
The exact set of ``necessary and sufficient'' conditions needed to provide fast transport of relativistic electrons along filaments is not clear. Given that the gas motions in the ICM are predominantly subsonic, we do not expect $\beta \sim 1$ in the bulk of the gas, although in some special conditions, e.g., a merger of two clusters and/or strong and sustained shear, patches of low $\beta \sim 1$ might be formed, driving $\varv_A$ closer to $c_{s,t}$. On the premise that velocities much higher than $c_s$ are required,  the mass density in the filaments has to be much lower than that of the ICM. In this case, AGN-produced plasma becomes the most viable candidate (as we also argue later in Sect.~\ref{s:discussion}). What fraction of volume could, in principle, be filled with such plasma? If we consider a typical cool-core cluster with the cooling radius $r_{\rm cool}\sim (0.1-0.2)\, R_{500c}$, i.e. $\sim 100-200\,{\rm kpc}$, the amount of energy supplied by the central AGN during the Hubble time will be of the order of the ICM thermal energy within the cooling radius. Assuming that all this energy is supplied in a non-thermal form and all ICM heating (AGN Feedback) is in fact set by “mechanical” interactions with non-thermal plasma \citep{2000A&A...356..788C}, i.e., effectively via adiabatic losses, the fraction of volume filled by non-thermal plasma within    $R_{500c}$  is between 2 and 10\%. This is, of course, a very crude estimate, but it shows that this fraction can be substantial. A much more difficult question is what happens to this non-thermal plasma? For instance, will it eventually be dispersed and mixed with the thermal ICM? Observations suggest that at least in the cores of clusters, there are “filament-like” structures that are a few 100 kpc long, which suggests a lifetime of more than 100~Myr (for velocities $\sim c_{s,t}$). Whether they survive longer is an open question; we further comment on the implications of this possibility in Sect.~\ref{s:discussion}.

\section{Examples of observational constraints on electron propagation velocities in visible filamentary structures}
\label{s:examples}
In this section, we provide examples where very high transport velocities for the radiating relativistic electrons provide a plausible explanation for the observed spectral structures.   These include very different astrophysical situations, all associated with galaxy clusters --  tailed radio galaxies and peripheral relics associated with merging clusters.  In each case, alternate explanations are also possible, such as distributed acceleration, but here our emphasis is on presenting the attractive possibility that fast electron transport is occurring.

Using Eq.~(\ref{e:tmax}), if the travel distance, $d$, of the electrons is measured, and the amount of radiative ageing over that distance is characterized by a quantity $t_{\rm obs}$, calculated from the change in the cutoff frequency in the spectrum, then a minimum electron velocity $\varv_{\rm min}$ can be derived.  To do this, we assume that the time of observation, $t_{\rm obs}$, is equal to the maximum lifetime, $t_{\rm max}$, for the observing frequency.   This yields a lower limit for the transport velocity of the electrons,  $\varv \gtrsim\varv_{\rm min}= d / t_{\rm max}$.  

\begin{figure}
\centering
\includegraphics[width=0.99\columnwidth,trim=0.1cm 0.cm 0.1cm 1.cm,clip]{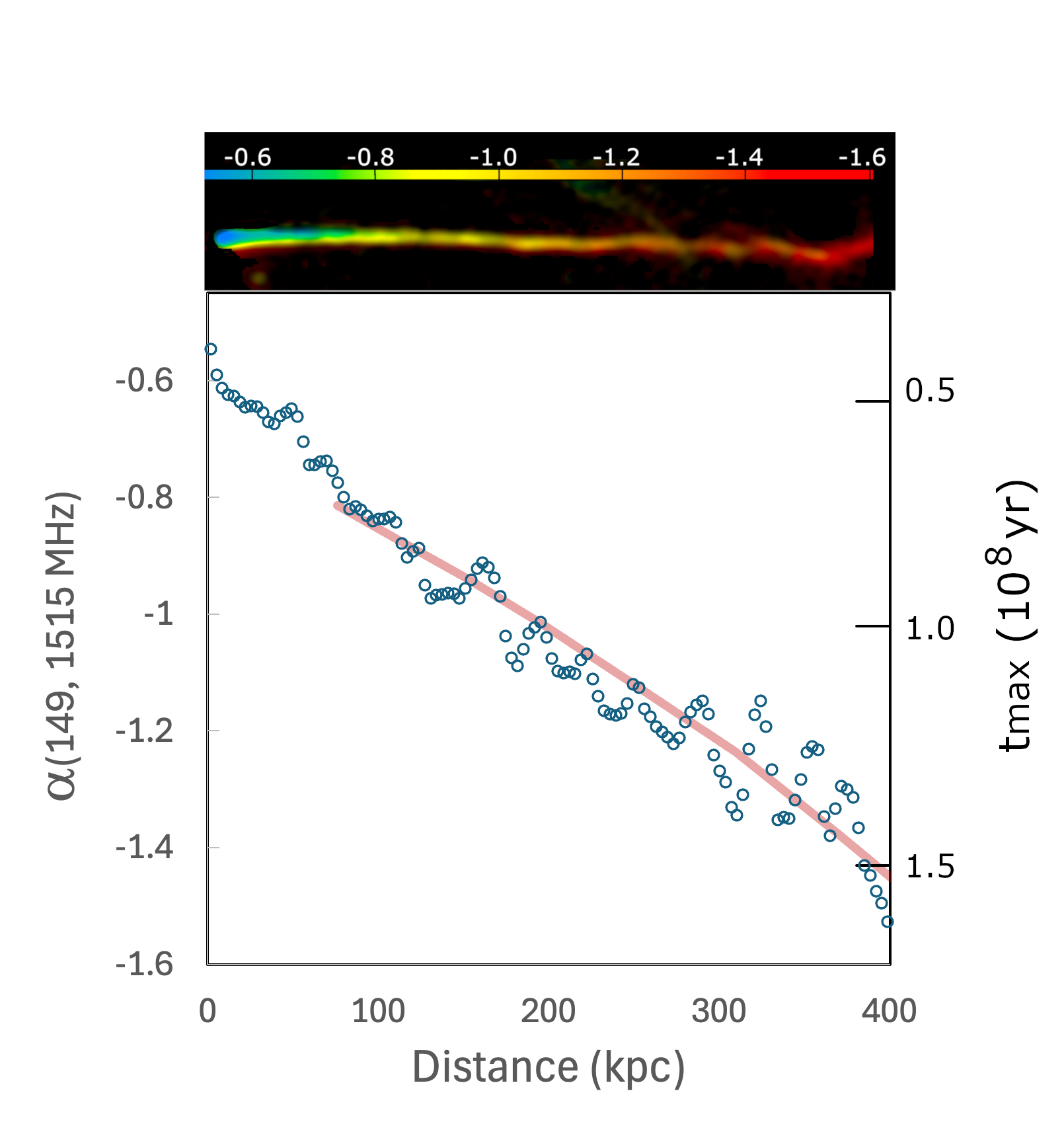}
\caption{  Top: Map of Tail C in Abell~2256 at 6.8\arcsec resolution at 149~MHz, (LOFAR reference), color-coded by its spectral index to 1.5~GHz \citep{2014ApJ...794...24O}. Bottom: Observed spectral index between 149 and 1515~MHz, and the corresponding maximum age for relativistic electrons at each location, based on the cutoff frequency calculated from the spectral index, as described in the text. The red line corresponds to an electron transport velocity of 4500~${\rm km\,s^{-1}}$.}
\label{fig:TailC}
\end{figure}

\subsection{Tailed radio galaxies}
We now examine two tailed radio galaxies, where the basic model assumes that the relativistic particles are (almost) at rest in the ICM. The jets emerge approximately transverse to the direction of motion of the host through the ICM, and the jets are swept back and quickly come approximately to rest\footnote{The jets can retain some of their initial velocity downstream, \citep[see, e.g., Fig. 5 in][]{2019ApJ...884...12O}, which, if known,  could make a small change to our calculations.}.   We look at two cases, Tail C in Abell 2256 (at $z\approx 0.058$) \citep{2014ApJ...794...24O, 2024A&A...688A.175O} and the GReET source in Abell 1033 \citep{2017SciA....3E1634D}. 

Tail C is a 700+ kpc very narrow tail in the cluster Abell 2256 \citep{1994ApJ...436..654R, 2014ApJ...794...24O, 2003AJ....125.2393M}. The host is SDSS 17330.04+783955.4 (WISEA J170330.04+783954.3), with a radial velocity of $17558\pm36\,{\rm km\,s^{-1}}$, nearly at rest along the line of sight with respect to the $17490\pm 74\,{\rm km\,s^{-1}}$ systemic cluster velocity. The cluster velocity dispersion is $1269^{+56}_{-49}\,{\rm km\,s^{-1}}$ \citep{2003AJ....125.2393M}.  Tail C's straightness and very long length have raised questions about how the host dynamics, or ongoing particle acceleration, could lead to the observations. Here, we re-examine the spectral behavior in the context of possible fast electron transport. 

The high-resolution observations shown in Fig.~14 of \citet{2014ApJ...794...24O} reveal that the structure is unresolved transversely in the first kpc downstream, and then shows parallel strands beyond $\sim$30~kpc from the host. These strands are blended into a single long, thin structure at the 6.8\arcsec resolution used here.  The presence of regions with parallel strands could be due to two very closely spaced tails or jet edge brightening, with Tail C  actually being a unique, extreme, one-sided, high-velocity jet. For the analysis presented here, we do not explore this option and treat the source as either a single, left-behind tail or two blended parallel left-behind tails.  

In Figure \ref{fig:TailC}, we plot the spectral index at a resolution of  $6.8''$ between 149~MHz (using a LOFAR map courtesy of R. van Weeren; see \citealt{2024A&A...688A.175O}) and 1515~MHz, (using a VLA map as presented in \citealt{2014ApJ...794...24O}).  From this, we calculate the local cutoff frequency, $\nu_c$, and thus $t_{\rm max}$ at each location, assuming an exponentially cut-off spectrum with a low-frequency index of $-0.5$, as observed near the host, and setting $B_1 = 2.26~{\rm \mu G}$ (using  Eq.~(\ref{e:b1}) and the CMB energy density at the cluster redshift).  The spectral index drops quickly ($\lesssim {\rm kpc}$) to approximately $-0.6$, and then more gently to about $-0.85$ at 70~kpc from the host. 
Further downstream, with some small but significant wiggles, the inferred $t_{\rm max}$ increases linearly with distance from the~AGN.  This is consistent with a constant velocity of the relativistic electrons (in the case of streaming) or of the host (in the case of flow being stopped in the ICM).

The relation between the slope of the spectral index and \mbox{distance}, converted to $\nu_c$~ and~ $t_{\rm max}$~ 
vs. distance,  yields an inferred relativistic electron velocity $\varv_{\rm min}\sim4500\,{\rm km\,s^{-1}}$ due to radiative ageing alone.  This represents a minimum velocity because, if the velocity were lower, or if $B \neq B_1$, then much more ageing would have occurred.  The velocity $\varv_{\rm min}$ would have to be even larger than derived above if there were also adiabatic losses, or if the magnetic field weakened downstream.  The brightness of the tail as a function of distance is approximately as expected from the spectral steepening alone. It is thus likely that the magnetic field is approximately constant in these first 400~kpc, so any weakening downstream does not contribute significantly to the change in spectral index.

If the tail were simply plasma at rest with the ICM, left behind after the passage of the host through this region, then the velocity of the host in the plane of the sky would have to be $\geq \varv_{\rm min}$.  Since  $\varv_{\rm min}$ is approximately 3.5 times higher than the cluster velocity dispersion, such a large projected velocity seems unlikely, although not impossible.   Fast streaming of the relativistic electrons down the tail, as proposed here, offers a simple way of achieving the required minimum velocity and reducing radiative losses.  Alternatives include a supersonic host velocity,  the long-standing idea of in situ acceleration of the relativistic particles throughout the tail \citep{1982ApJ...254..472E,2025MNRAS.539.3697D}, or that Tail~C actually is a unique one-sided jet with a bulk flow of $\sim 4500 \, {\rm km\,s^{-1}}$.  

Another source with a remarkably constant spectral index in its downstream tail is found in the southern portion of Abell~1033  \citep[hereinafter GReET,][]{2017SciA....3E1634D, 2022A&A...666A...3E}.  The source is a ``narrow-angle-tail'' (NAT)\footnote{GReET mistakenly characterized it as a ``wide-angle-tail'' \citep{1976ApJ...205L...1O}; the origins of this type of error are described in \citet{2021Galax...9...85R}.} similar, e.g., to NGC~1265.  Far behind the host AGN, a larger than $200\,{\rm kpc}$ portion of the tail is displaced transversely, which is attributed to bulk motions in the medium, perhaps behind a weak shock. Such displacements are indeed seen in the simulations of \citet{2019ApJ...887...26O},  who introduce a shock moving perpendicular to a NAT tail.   

From multi-frequency observations and color-color diagrams \citep{1993ApJ...407..549K}, \citet{2017SciA....3E1634D} and \cite{2022A&A...666A...3E} measure GReET's injection spectral index to be -0.65. The spectra then steepen downstream to -2 (between 54 and 323~MHz) and -4 (between 144 and 323~MHz). Beyond 400~kpc, the spectral steepenings should continue even more dramatically, with continuing radiative losses, but this is not observed. \cite{2017SciA....3E1634D} invoke ``gentle reacceleration'' by turbulence in the displaced portion of the tail,  in order to maintain a relatively constant, but steep spectrum. As noted by \citet{2017SciA....3E1634D} this requires ``a very special geometrical and physical configuration.''  
We estimate that local fluctuations in Mach number, if DSA were responsible, would need to be less than a few percent, which seems unlikely. Whether or not the turbulence has uniform enough properties over 100s of kpc to maintain the spectral shape is something to explore. 
Here, we investigate the possibility that, instead,  large electron velocities in this portion of the tail minimize the spectral steepening.

\begin{figure}
\centering
\includegraphics[width=0.9\columnwidth]{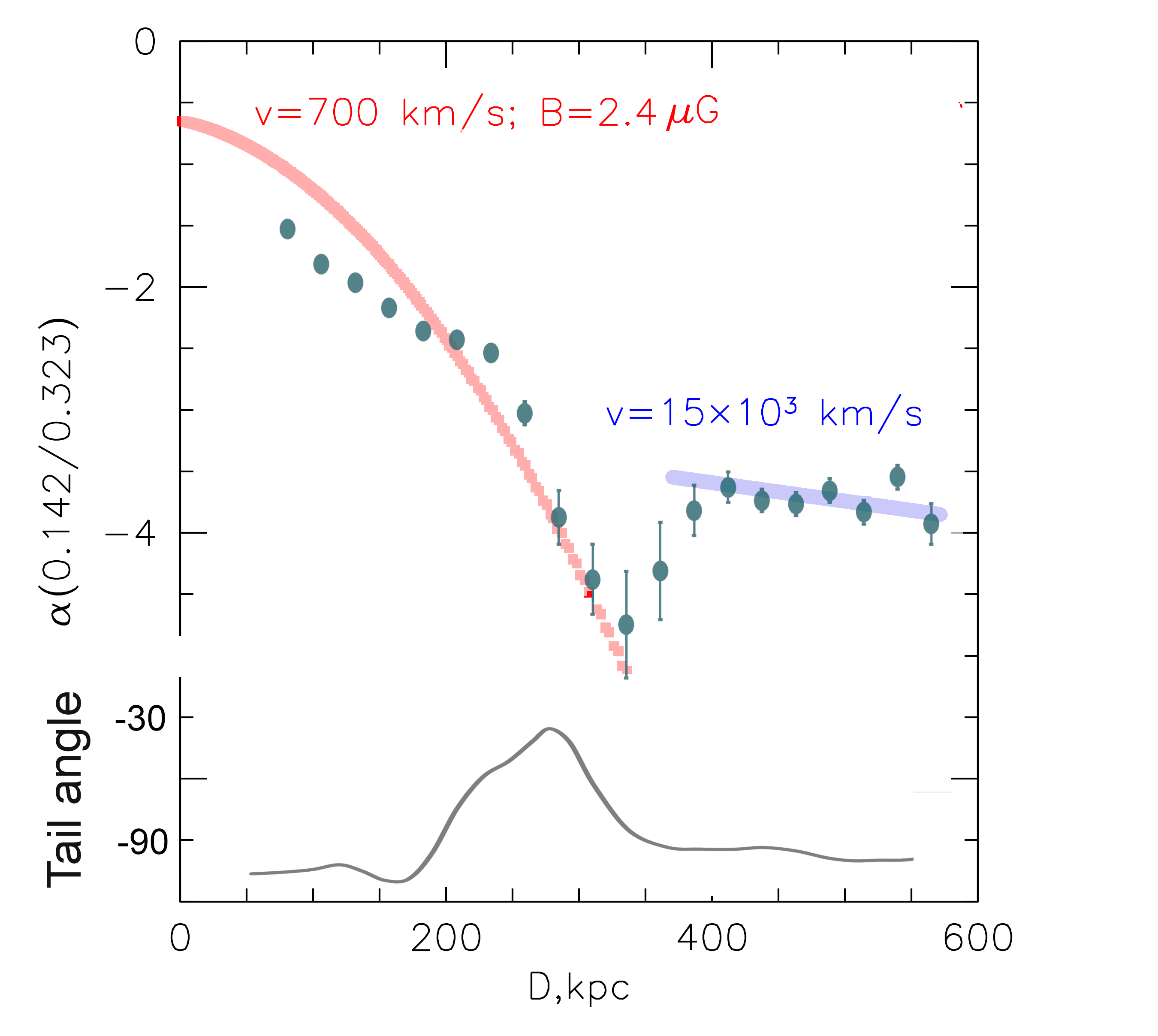}
\caption{The spectral index of the GReET tail in A1033 tail as a function of distance from the host, from Fig.~3 in  \cite{2017SciA....3E1634D}. The red points show the expected steepening of an $\alpha_0= -0.65$ spectrum with a low velocity for the radiating plasma,  while the very modest amount of steepening in the final portion of the tail is consistent with a much higher transport velocity of 15,000~km/s, as shown in blue.  The bottom plot, in solid black, shows the inferred direction of the host's motion at each position in the tail,  assuming the tail is left behind material. The variation in direction around 400~kpc is assumed to arise from a transverse flow, not from a change in the host velocity. } 
\label{f:a1033_lr}
\end{figure}

We performed a similar analysis as for Tail~C above.  Figure \ref{f:a1033_lr} shows a plot of the spectral index between 144 and 323~MHz, taken directly from Fig.3 of  \cite{2017SciA....3E1634D}.  Behind the host, the spectral index steepens from $\sim-1.5$ to $\sim-4.5$ in the first 350~kpc.  With a constant magnetic field $B_1 = 2.4 ~{\rm \mu G}$ (thus minimizing losses with $B_{\rm CMB}/\sqrt{3}$)  and a velocity of 700~${\rm km\,s^{-1}}$, the observed steepening with distance would be reproduced. But after flattening to $\sim-3.75$ (discussed further below), the spectral index stays relatively constant for the final 200~kpc;  it is this last section of the tail that we consider.  Here, the maximum amount of steepening consistent with the data is from $-3.75$ to $-4$.  Since we are in such a curved part of the spectrum, from the initial $-0.65$ power law,  this corresponds to only a very small change in $\nu_c$, from 72~MHz to 67~MHz.  Again using  $B_1 = 2.4~{\rm \mu G}$, this corresponds to a change in $t_{\rm max}$ from  375~Myr to  only  389~Myr, giving $\varv_{\rm min}\sim15,000\,{\rm km\,s^{-1}}$.  Since this velocity is much higher than any plausible dynamical velocity in the system, fast electron transport becomes an attractive possible cause.

The fast transport process alone does not explain the flattening of the spectral index between $\sim$375 and 400~kpc from the host. However, the tail is fainter in the steepest region, so that the magnetic field could be somewhat weaker there, then strengthening again as the brightness increases and the spectra flatten.  Another interesting possibility is that the flattening arises not from a change in magnetic field, but from the bending of the tail along the line of sight.  In Figure  \ref{f:a1033_lr}, we plot the inferred direction of the host velocity in the plane of the sky; its initial trajectory is at $\sim-105^{\circ}$,  then it curves by $\sim60^{\circ}$ as it is transversely displaced approaching  400~kpc, until returning to $\sim-95^{\circ}$, close to its original direction.  If this dramatic change in curvature were accompanied by only a $30^{\circ}$ change in direction along the line of sight, this would cause the observed spectrum to flatten from -4.79 to -4.3, as observed.  This occurs because the observed cutoff frequency is, in more detail,  given by $\nu_c\propto \gamma_c^2 B \sin \theta$, where $\gamma_c$ is the electron cutoff energy and $\theta$ is the angle to the line of sight.
Thus, in the case of a strong ordered magnetic field, at a fixed observing frequency, the observed spectral indices from a curved spectrum depend on the angle of the field.  The analysis above does not constitute proof that fast electron transport and possible jet bending are the cause of GReET's spectral behavior; it does show that it is plausible.

A close analog to the spectral trends in GReET 
was recognized during the revision process, viz. the MysTail in Abell 3266 \citep{2021Galax...9...81R, 2022A&A...657A..56K, 2022MNRAS.515.1871R}. Like GReEt, it is tangentially oriented in the periphery of the cluster X-ray emission.  Its radio spectra steepen quickly with distance from the host for the initial 200+~kpc, followed by a fainter region of another 200+~kpc with little or no further spectral steepening (see Fig. 5 in \citealt{2021Galax...9...81R} and Fig. 13 in \citealt{2022MNRAS.515.1871R}). Embedded in the tail are at least three prominent filaments connecting two diffuse patches of emission \citep[Fig. 7 in ][]{2021Galax...9...81R}; other striking filaments are shown in their Fig. 10, connecting with the long tail in Abell 1314 \citep{2021A&A...651A.115V}.

\subsection{Filaments in cluster relics}
\begin{figure}
\centering
\includegraphics[clip,trim=0.1cm 0cm 0.1cm 0cm,width=0.8\columnwidth]{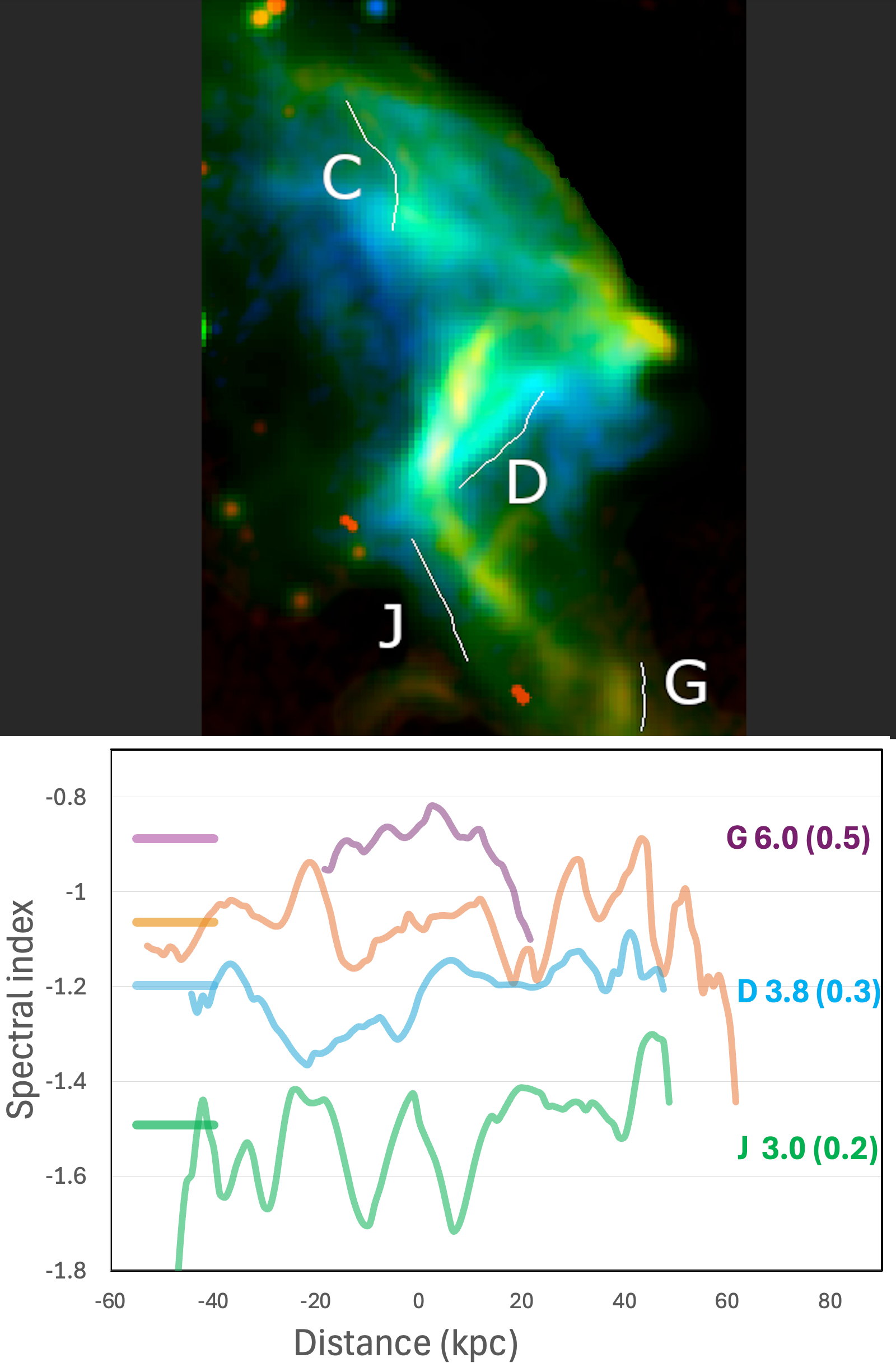}
\caption{{Top: NW relic of Abell 3667 from \cite{2017SciA....3E1634D}.  Colors indicate the spectral indices, and different filaments show different characteristic indices. Bottom:  spectral indices along five of the filaments, showing how the variations along them are significantly smaller than the differences between them.  Numbers on the right indicate the equivalent Mach numbers (and their scatter) if these indices were due to DSA. Errors per measured point are just slightly smaller than the variations, so the intrinsic variations are significantly smaller than observed. 
}}
\label{fig:A3667}
\end{figure}

In this section, we look at the northwest relic of Abell 3667, based on the data presented in \cite{2022A&A...659A.146D}.  The relic is highly filamentary, with a mixture of short segments on scales of $10^2$~kpc, and a number of much longer filaments, on Mpc scales, comparable to the size of the entire relic.  Our Figure \ref{fig:A3667} is a zoomed-in portion of their Fig. 14, revealing the wide range of spectral indices within the relic. One striking aspect of these spectral variations is that they appear well-correlated with the filamentary structures themselves -- each filament has its own characteristic spectral index, along with some variations. 

To look at this more quantitatively, we measured the spectral indices along the ridge line of several of these filaments, as shown in the bottom of Figure \ref{fig:A3667}.  The average index of each of these filaments is significantly different than the others.  The intrinsic spectral variations along these filaments are likely significantly smaller than observed, since there is an extensive network of overlapping structures and underlying diffuse emission of comparable brightness. 

A complicated shock substructure is seen in many numerical simulations \citep[e.g.,][]{2008MNRAS.391.1511H,2011ApJ...735...96S,2017MNRAS.464.4448W,2017MNRAS.470..240N,2019ApJ...883..138R,2021MNRAS.500..795D,2022MNRAS.509.1160I,2023MNRAS.526.4234S,2023ApJ...957L..16B,2025ApJ...978..122L}.  
In such a complex environment, the question is how to maintain a relatively constant spectral index along each filament.  Their shapes and structures make it unlikely that the local Mach number remains strictly constant along each filament.  The variations along the segment D, for example, would require Mach numbers constant to within $\pm8\%$. MHD simulations of relics show a wide range of Mach numbers within each relic \citep[e.g.,][]{2023ApJ...957L..16B,2025ApJ...978..122L}, so the broad range of values observed here between different filaments is expected;  what is surprising is the coherence of spectral indices over long distances within individual filaments.

One explanation for the observed spectral uniformity of filaments in A3667 is that any local variations in Mach number/particle acceleration have been homogenized due to fast transport along the filament.  As long as the homogenized spectra are close to a power law over just a factor of a few in frequency, localized magnetic field variations will not lead to varying observed spectral indices.  In the absence of significant radiative losses around the frequency of observation, we expect this to be the case.  
Higher-resolution observations at a range of frequencies would allow the spectra and spectral curvature of filaments to be disentangled from the various confusing structures and provide better constraints on their spectral uniformity. 

Abell 2256 is another case where a prominent network of filaments fills the relic \citep[e.g.,][]{2006AJ....131.2900C}.  Using the data in \cite{2014ApJ...794...24O}, we attempted to trace the spectral variations along individual filaments, to see if they had distinct values.  The variations along filaments, however, were of the same order as the difference between filaments, and therefore do not provide evidence for fast electron transport.  However, since the filaments overlap with one another and are each embedded in more extended features, we could not conduct a clean test.  It would be useful to experiment with a range of spatial filtering and tomographic analyses to see whether individual structures with distinct spectral indices could be isolated.

\section{Discussion}
\label{s:discussion}

The above considerations suggest that in a non-thermal filament with a strong field, the energy spectrum of relativistic electrons can be largely independent of the distance from the initial source of these electrons.  A  condition for such spectral uniformity is that the propagation time of electrons along the filament is shorter than both the radiative cooling time and the filament expansion time    $\left ({d \ln V}/{d t}\right)^{-1}$, where $V$ is the filament volume. Given that the Alfvén velocity $\varv_{A,f}$ in the filament can be high, this appears to be the easiest scenario that allows for  ``synchronizing'' the electron populations emitting at $\lesssim$GHz frequencies over scales $\sim 0.1-1\,{\rm Mpc}$ in galaxy clusters.   We also presented several examples of visible filamentary structures in clusters of galaxies where this could be taking place.

\subsection{Formation of filaments}
How can filaments with such properties be formed in clusters? As mentioned in Sect.~\ref{s:survival}, one natural scenario is related to AGN-inflated bubbles in cool-core clusters. Observations in radio and X-ray bands \citep[e.g.,][]{1993MNRAS.264L..25B} show that these bubbles are bright in radio and dim in X-rays, suggesting that magnetic fields and relativistic particles are present, while the density of thermal plasma is significantly lower inside the bubbles. Even stronger limits come from observations of the SZ \citep{1972CoASP...4..173S}  signal from the largest resolved bubbles in MS~0735.6+7421 \citep{2019ApJ...871..195A,2022A&A...667L...6O}. The lack of the SZ decrement from these bubbles suggests that either they are devoid of thermal plasma or this plasma has a temperature higher than hundreds of keV. This makes AGN-inflated bubbles good candidates for plasma reservoirs with the required properties.  These light ``bubbles'' evolve under the action of buoyancy \citep[e.g.,][]{1973Natur.244...80G,2000A&A...356..788C} and move radially in cluster atmospheres down the pressure gradient and/or due to large-scale motions of ambient thermal plasma.   Buoyancy naturally produces subsonic motions \citep[e.g.,][]{2018MNRAS.478.4785Z} and can also stretch magnetic field lines that are anchored to the bubble and the gas in the core \citep[e.g.,][]{2013MNRAS.436..526C}. Once again, the subsonic velocity implies that an isolated bubble should suffer from radiative (and adiabatic) losses. However, once a filament is formed, the signs of both types of losses along the filament can diminish.
This scenario can be relevant to objects like M87, Nest200047, Abell~194, and other clusters where a central AGN is present. Yet another scenario, involving a central AGN, was considered in \cite{2022ApJ...934...49G} in application to the A3562 cluster. In this scenario, tangential flows along the contact discontinuity in the cluster core create an elongated filament. Once again, either reacceleration or fast propagation is needed to keep the radio spectrum uniform.

Another possibility is that filaments originate in the flows from one or more cluster non-central AGNs.  For example, in tailed galaxies, the AGN is still responsible for producing the initial volume of magnetized plasma, but the motion of the galaxy relative to the ICM produces a long ``channel'' with non-thermal plasma. Similarly to the original model of \cite{1973A&A....26..423J}, the (strong) field becomes ``straight'' in the far tail, while in the vicinity of the source, it might remain turbulent and tangled. 

A pre-existing population of filaments will also affect the appearance of radio relics.
For relics, the presence of shocks is a strong argument for an acceleration (or re-acceleration) scenario. 
As in \cite{2023A&A...670A.156C}, we argue that when accelerated (or re-accelerated) leptons appear in a preexisting filamentary structure with an ordered strong field, these leptons can quickly propagate along the entire structure. Once again, an AGN might be the initial source of these low-beta structures, further stretched and bent by the gas motions as, e.g., in simulations of \cite{2021ApJ...914...73Z}.  

On completely different physical scales, the Galactic Center NTFs could arise from stretching and ordering of the field by a global outflow from the central region \citep[e.g.,][]{2019Natur.573..235H,2019Natur.567..347P}. This demonstrates that conditions favorable for fast propagation are not endemic to clusters.

\subsection{Testable predictions of the model}
In this section, we discuss some possible testable predictions of the model discussed here. These include magnetic field topology, radio spectral properties/trends, and, possibly, IC emission.

\vspace{0.2cm}

\subsubsection{Magnetic field orientation, degree of polarization, Faraday rotation}

First, a strong field aligned with the filament implies that the observed synchrotron emission should be strongly polarized. Indeed, for some filaments with measured polarization, the orientation of the electric field vector is perpendicular to the filament, and the degree of polarization is large \citep[e.g.,][]{1984Natur.310..557Y,2019ApJ...884..170P,2022ApJ...935..168R}. A patch of the Radio Arc near the PWN candidate G0.13-0.11 features strong polarization ($57\pm18$\%) even in the X-ray band \citep{2024A&A...686A..14C}, where Faraday rotation is negligible. This level of polarization, albeit with large uncertainties, is consistent with a highly ordered field.  The same is true for tailed radio galaxies, e.g., the classic case of 3C129 \citep{1973A&A....26..413M, 2016MNRAS.461.3516M}, as well as newer examples \citep{2021MNRAS.508.5326M}.

Since this model assumes a low density of electrons inside filaments where synchrotron radiation is produced, and the fields are highly ordered, no internal Faraday rotation is expected, and the fractional polarizations of filaments should be high, as is generally observed.  In the one case where low-$\beta$ conditions have been established from the presence of an X-ray cavity (the ``E-fils'' in Abell 194, \citealt{2022ApJ...935..168R}), the fractional polarizations are extremely high (1$\pm$0.25). In this case, the observed large coherent patterns in the Faraday rotation likely arise in the local surrounding medium. In Abell 194, the Faraday-corrected magnetic field angles trace well the local orientation of the filaments. Observations of magnetic field directions that were not along the filaments in some objects would also invalidate the model for them. Unfortunately, current X-ray instruments are not capable of detecting an excess of thermal emission due to filaments, because of limited angular resolution and sensitivity.

In the absence of depolarization, highly polarized filaments provide other opportunities for studying the intracluster medium.  Faraday rotation patterns along a sufficiently long filament have distinct advantages in measuring the correlation length of the magnetic field in the ICM. It can also provide clues to the 3D structure of the filament, as suggested by \citet{2022ApJ...935..168R}. An interesting additional consequence of this model, once proven,  is that since the magnetic field would be parallel to the filament, measurements of the polarization angle at a single frequency would be sufficient to measure the Faraday rotation. 

\vspace{0.25cm}
\subsubsection{Shape of the synchrotron spectra}

Second, if all energy losses of particles are indeed minimal, any existing break in the particle distribution does not change along the filament. This should be reflected in the shape of the observed synchrotron spectra. The persistence of the spectral shape can be studied through the use of color-color diagrams \citep{1993ApJ...407..549K}.  In addition, if there is an independent way to connect the local magnetic field strength in the filament to the surrounding thermal pressure, then one can make predictions about the correlation between the observed filament brightness and the observed $\nu_b$.  An object such as the Nest20047 group \citep{2025A&A...696A.239B} may present such an opportunity. 
The condition of minimal losses for magnetically supported structures in cluster cores implies that the characteristic frequency of synchrotron radiation 
should scale as $\nu_b\propto \gamma^2 B_f\propto P_t^{1/2}$. This is in stark contrast with the maximal-losses scenario, when adiabatic and radiative losses ``cooperate'' on the background of slowly moving buoyant bubbles (see Fig.~\ref{f:age}). The presence of emission regions with ``young'' spectra, some 100~kpc from the core, suggests that either re-acceleration or fast propagation mechanisms are involved. Discriminating between different scenarios might be non-trivial. Once again, a high degree of polarization might be particularly relevant for the fast-propagation scenario.

Yet another prediction relies on the possibility that leptons can go back and forth along a filament before they age. As discussed in Sect.~\ref{s:radiative}, a possible configuration might involve a transition to a high-$\beta$ phase at the end of the filament, where the scattering rate is large \citep[e.g.,][]{2024MNRAS.532.2098E,2025NatAs...9..438R}. Alternatively, the filament might be magnetically isolated from the ambient plasma so that the field lines are tangled at the end of the filament.   
If the leptons are added to the filament at different times, we should see a mixture of different ages at any given position. This implies that we should see not a simple single-age spectrum characterized by one break frequency but a gently curved shape. In this case, models that consider a distribution of ages \citep[e.g.,][]{1994A&A...285...27K} might provide a better fit to the data than models that assume a single episode of particle acceleration.   

In the model considered here, large-scale geometry is set by either buoyancy or the motion of an AGN relative to the ICM. At the same time, coherent radio structures are synchronized by the propagation of electrons along the field lines, while the cross-field diffusion is strongly suppressed. As a result, one can expect to see ``bundles of threads'' in the radio images with sufficient sensitivity and angular resolution. This is indeed observed for NTFs \citep[e.g.,][]{2019ApJ...884..170P} and the filaments in clusters \citep[e.g.,][]{2022ApJ...935..168R,2025A&A...696A.239B,2024A&A...692A..12V,2025A&A...699A.229D}.  However, the filaments locally are essentially ``independent'', i.e., each of them is ``synchronized'' along its length but not necessarily with other filaments. As a result, individual filaments might be unresolved or too faint to be seen.

Even if the electron spectrum is homogenized along the filament and the pitch-angle distribution of electrons is isotropic, one can expect spectral variations due to the changes of the angle $\theta$ between the filament ($=$ magnetic field) and the line of sight.   In the ordered field, the intensity and critical frequency both depend on $B\sin\theta$. 
For a narrow filament, the increase of the line-of-sight length compensates for the decrease of intensity for small values of $\theta$, but the critical frequency dependence on $\theta$ remains.
Therefore, one can expect to find a correlation between the brightness and hardness of the spectrum, provided the cutoff frequency is not much higher than the observed range. Such a correlation will have the same signature as the variations of the age, except it reflects only the orientation of the filament. In this case, the dependence on the viewing angle might be non-monotonic along the filament, and a correlation or anticorrelation of the flux and slope is possible. We speculate that in some extended radio sources \citep[e.g.,][]{2017SciA....3E1634D}, this effect can explain the sudden hardening of the synchrotron emission further away from the primary source of electrons.

On a more speculative side, if the filaments are formed by large-scale motions stretching an initially compact region, magnetic field lines can reverse direction across filaments (see, e.g., \cite{1973A&A....26..423J} for radio tails or \citealt{2013MNRAS.436..526C} for radio bubbles). Such structures might have weaker fields in the center, leading to edge-brightened ``hollow-tubes'' in radio emission. It is interesting to note that several ``double-stranded''  filaments are indeed seen in clusters (e.g., \citealt{2021NatAs...5.1261B}, see also our Fig.~\ref{f:img}). A reversal in magnetic-field direction might be detectable by transverse gradients in Faraday rotation \citep{2024MNRAS.535.2115R}, when there is sufficient spatial resolution.

\vspace{0.25cm}
\subsubsection{Long-living filaments and IC emission}

If the filaments can survive without mixing with thermal gas for a long time (longer than the synchrotron cooling time of electrons emitting in the observable band), the ICM might be  “infested” with filaments occupying a small volume fraction but present everywhere. In principle, the presence of such (long-living and most of the time invisible) structures might affect dynamical or transport properties of the bulk ICM plasma.

This might also affect the propagation of cosmic rays on spatial scales set by the sizes/morphology of filaments embedded in the ICM. If relativistic protons make a significant contribution to the non-thermal pressure but are confined to filaments, no gamma-rays associated with $\pi^0$ are expected \citep[e.g.,][]{2017MNRAS.470.3388P}. On the contrary, the IC emission associated with relativistic leptons might be enhanced if these leptons are confined to non-thermal filaments with a small density of thermal electrons, effectively suppressing the Coulomb losses for the low-energy part of the particle spectrum. Indeed, in the majority of models \citep[e.g.,][]{1999ApJ...520..529S,2001ApJ...557..560P,2025arXiv250718712H}, the Coulomb losses are calculated assuming a background density of electrons $n_e\sim 10^{-3}-10^{-2}\,{\rm cm^{-3}}$. In the “filaments” model, the Coulomb losses are absent (or suppressed), and the peak of the aged distribution of relativistic electrons is either absent or shifted towards lower Lorentz factors. Since the Coulomb losses are suppressed, the non-thermal bremsstrahlung is absent, too. The synchrotron losses are instead high (in cluster cores, if magnetic field energy density is close to ICM pressure), further shifting the peak Lorentz factor to lower values. Therefore, in the no-mixing scenario, IC emission in the optical/UV band is possible, which might be relevant to the reported UV excess from clusters \citep[e.g.,][]{1996Sci...274.1335L}, especially given that the (thermal and photonionization-induced) soft X-ray emission from the IGM predicted by numerical simulations is orders of magnitude fainter and has different radial dependence \citep[][]{2023MNRAS.523.1209C}. However, to explain the reported UV luminosity relative to the observed radio luminosity, an aged electron population is required, as in several models discussed in \cite{1999ApJ...520..529S}, except that the break in the spectrum due to Coulomb losses can now be shifted towards lower Lorentz factors (or frequencies in the observed spectra).

\subsection{Are such objects rare? And the need for more observations}

Finally, we address the question of why clear cases for lossless transport appear so rarely.  In short, the main reason is that we have not yet looked for them, and there are few existing broad-band analyses spanning large variations in spectral index.  The example of MysTail \citep[][see Fig. 1]{2021Galax...9...81R,2022MNRAS.515.1871R} a case in point;  although this work was done by one of the authors, we did not realize that it provided an even more interesting case than the GReET source, because there is no alternative explanation, such as an impinging shock, for the $>200$~kpc long near constant electron cutoff energy.  It is important that observers actively look for additional cases where the initial losses in an inferred flow appear to have halted or dramatically slowed.  More opportunities for this are rapidly emerging from sensitive observations with LOFAR, MeerKAT, and ASKAP.

A more subtle, but no less important consideration is that analyses in the literature of presumed radiative losses commonly fail to account for the steepening of spectra that will occur when an electron population with a cutoff energy is viewed in progressively lower magnetic fields.  This is exacerbated when there are adiabatic losses as well, such as due to expansion down a tail or lobe.  Only after accounting for those spectral steepening effects can we ascertain whether additional radiative losses have or have not occurred.  This is not a straightforward process, since the observables are degenerate with respect to the underlying physical parameters of cutoff energy, magnetic field strength, and relativistic electron number density.  A prescription for isolating the physical parameters, in favorable cases, is presented in \citet{1994ApJ...426..116K}.

At an even more basic level, our ability to detect filamentary structures is heavily dependent on observational sensitivity, resolution, dynamic range, the filling/covering factors for filamentary structures, and their spectral characteristics.  Following the history of how filamentary structures have been revealed so far, with improved observational techniques, we expect that they will be found in many more situations in the future. At the next level, statistical studies of their spectral index variations, polarization fractions, and magnetic field orientations, substructures, etc., will be needed to constrain the underlying physics.

\vspace{0.25cm}

\section{Conclusions}
We argue that there is empirical evidence that low plasma beta ($\beta\lesssim 1$) non-thermal filaments exist in galaxy clusters and other environments. Such filaments can effectively minimize energy losses of relativistic electrons propagating along them and give rise to extended structures in the synchrotron emission with similar spectral properties (e.g., radio relics) and/or an effective break frequency slowly changing 
with distance from cluster cores (reflecting radial variation of the magnetic field rather than changes in the particles' energy).  Such filaments in cluster cores may contain a mixture of electrons with different ages, resulting in a gently curved spectrum. The magnetic field is expected to be laminar and aligned with a filament direction; therefore, synchrotron emission should be strongly polarized perpendicular to the filament. This model predicts a positive correlation between the intensity and the spectrum hardness associated with variations of the filament orientation relative to the line of sight. For instance, a long but bent (relative to the line of sight) filament or tail can appear brighter and ``younger'' or fainter and ``older'' multiple times along its projected length, even if the pitch angles are isotropic, the magnetic field is constant, and the electron population is the same.  Overall, the filamentary plasma should have very different correlation lengths of all its properties along and across the filaments. The individual filaments can be very narrow and observationally unresolved or too faint to be detected, but still provide a channel for electron transport. It is important to search both the literature and, in particular, upcoming sensitive multifrequency observations, for cases where lossless transport may be significant.

\begin{acknowledgements}
We are grateful to our reviewer for the insightful comments.
IK acknowledges support by the COMPLEX project from the European Research Council (ERC) under the European Union’s Horizon 2020 research and innovation program grant agreement ERC-2019-AdG 882679. The work of AAS was supported in part by the UK STFC grant ST/W000903/1) and by the Simons Foundation via a Simons Investigator Award. MB acknowledges financial support from Next Generation EU funds within the National Recovery and Resilience Plan (PNRR), Mission 4 – Education and Research, Component 2 – From Research to Business (M4C2), Investment Line 3.1 – Strengthening and creation of Research Infrastructures, Project IR0000034 – “STILES – Strengthening the Italian Leadership in ELT and SKA”, from INAF under the Large GO 2024 funding scheme (project "MeerKAT and Euclid Team up: Exploring the galaxy-halo connection at cosmic noon”), the Large Grant 2022 funding scheme (project "MeerKAT and LOFAR Team up: a Unique Radio Window on Galaxy/AGN co-Evolution") and the Mini Grant 2023 funding scheme (project ‘Low radio frequencies as a probe of AGN jet feedback at low and high redshift’). We appreciate the advice on relativistic particle acceleration from Dongsu Ryu, Franco Vazza, and Tom Jones.

\end{acknowledgements}

\bibliographystyle{aa}
\bibliography{ref}

\label{lastpage}
\end{document}